\documentclass[aip,rsi,twocolumn,10pt]{revtex4-1}
\usepackage{graphicx}
\usepackage{amsmath}
\usepackage{amssymb}
\usepackage[latin1]{inputenc}
\begin{document}
\title{Creation of Quantum-Degenerate Gases of Ytterbium in a Compact 2D-/3D-MOT Setup}
\author{S{\"o}ren D{\"o}rscher}
\thanks{S. D{\"o}rscher and A. Thobe contributed equally to this work.}
\author{Alexander Thobe}
\thanks{S. D{\"o}rscher and A. Thobe contributed equally to this work.}
\author{Bastian Hundt}
\author{Andr\'{e} Kochanke}
\author{Rodolphe Le Targat}
\author{Patrick Windpassinger}
\author{Christoph Becker}
\author{Klaus Sengstock}
\email{klaus.sengstock@physnet.uni-hamburg.de}
\affiliation{Institut f{\"u}r Laserphysik, Zentrum f{\"u}r Optische Quantentechnologien, Universit{\"a}t Hamburg, Hamburg, 22761, Germany}
\date{\today}
\begin{abstract}
We report on the first experimental setup based on a 2D-/3D-MOT scheme to create both Bose-Einstein condensates and degenerate Fermi gases of several ytterbium isotopes. Our setup does not require a Zeeman slower and offers the flexibility to simultaneously produce ultracold samples of other atomic species. Furthermore, the extraordinary optical access favors future experiments in optical lattices. A 2D-MOT on the strong ${}^1S_0\rightarrow{}^1P_1$ transition captures ytterbium directly from a dispenser of atoms and loads a 3D-MOT on the narrow ${}^1S_0\rightarrow{}^3P_1$ intercombination transition. Subsequently, atoms are transferred to a crossed optical dipole trap and cooled evaporatively to quantum degeneracy.
\end{abstract}
\maketitle
\section{Introduction}
Owing to their unique properties, alkaline-earth-like atoms have become the subject of intense experimental and theoretical research in recent years. Ultracold gases of these two-electron atoms have been proposed for the realization and study of a variety of novel quantum systems including heavy fermion materials and the Kondo insulator\cite{Foss-Feig_2010A, Foss-Feig_2010B, Foss-Feig_2011}, SU(N)-symmetric Hamiltonians\cite{Gorshkov_2010}, quantum information processing\cite{Daley_2008, Gorshkov_2009} and artificial gauge fields\cite{Gerbier_2010, Dalibard_2011, Szirmai_2011}. Quantum degenerate gases have been reported by only a few groups for isotopes of ytterbium\cite{Takasu_2003, Fukuhara_2007A, Fukuhara_2007B, Fukuhara_2009, Taie_2010, Sugawa_2011, Hansen_2011} (Yb), strontium\cite{Stellmer_2009, deEscobar_2009, DeSalvo_2010, Tey_2010, Stellmer_2010}, and calcium\cite{Kraft_2009, Halder_2012}. However, all of these experiments rely on Zeeman slowers as the initial cooling stage.

From the early days of laser cooling\cite{Pritchard_1987} it is well known that Zeeman\cite{Phillips_1982} or chirp\cite{Ertmer_1985} slowing of a thermal beam enhances the loading rate of a magneto-optical trap (MOT) by many orders of magnitude as compared to loading from a background vapor directly. For alkali metals the concept of a 2D-MOT\cite{Dieckmann_1998} has been developed as a powerful alternative which results in a similar flux of cold atoms, but in a much more compact and versatile setup. Here, atoms are transversely cooled from a background vapor into a well-collimated beam of cold atoms. Because longitudinally slow atoms are cooled preferentially, the resulting velocity distribution peaks at several tens of meters per second allowing them to be captured by the 3D-MOT. Often, a ``pushing'' beam or moving molasses is used to enhance the yield of the cold atom source further. 2D-MOTs are extremely efficient, yet uniquely simple and highly tunable. In particular, a single setup can readily be used to cool multiple elements, whereas multispecies Zeeman slowers require sophisticated designs. 2D-MOTs are therefore inherently well suited for producing and studying ultracold mixtures.\cite{Ospelkaus_2006} Furthermore, high-precision spectroscopy experiments, which are prone to frequency shifts from black-body radiation, can benefit from the fact that there is no direct line of sight from the source of thermal atoms to the cold or ultracold sample.

The traditional concept of a 2D-MOT is limited to elements with sufficiently large vapor pressures near room temperature. Alkaline-earth-like elements, except for mercury\cite{Petersen_2008}, have high melting temperatures $T_l$ (e.g. $T_l = 1097\ \mathrm{K}$ for Yb) and low vapor pressures (e.g. less than $10^{-15}\ \mathrm{Pa}$ at $T\approx300\ \mathrm{K}$ for Yb{\cite{Habermann_1964}}). At room temperature this effectively prevents any 2D-MOT from being loaded from a background gas. An alternative concept has been demonstrated for lithium\cite{Tiecke_2009}, where the 2D-MOT is side-loaded from an oven instead of background vapor and high loading rates of up to $10^9\ \mathrm{s}^{-1}$ have been reported.

In this paper we report on the first successful implementation of a 2D-MOT loaded via a dispenser for Yb and the subsequent generation of quantum degenerate gases of both bosonic ${}^{174}\mathrm{Yb}$ and fermionic ${}^{173}\mathrm{Yb}$. Based on numerical simulations, we have designed a very compact ``glass-cell'' setup for a 2D- and 3D-MOT that allows extraordinarily large optical access for future experiments based on optical lattices.

\section{Experimental Concept}
\subsection{Atomic Properties of Yb}
\begin{figure}
	\centering
	\includegraphics{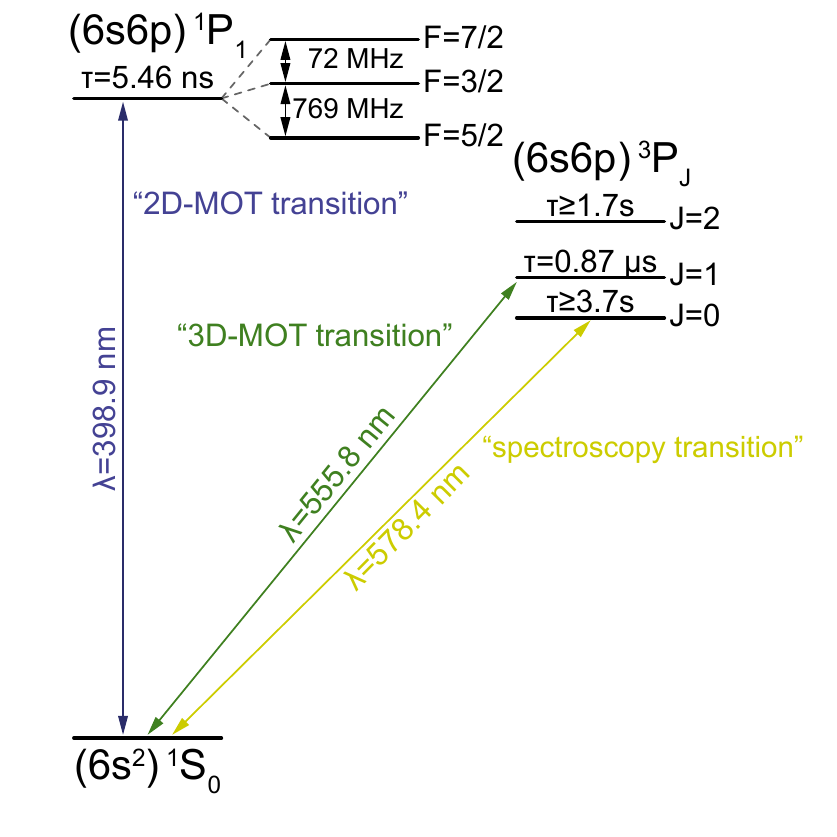}
	\caption{Partial Level Scheme of Yb. Only the lowest states in the triplet and singlet manifolds, which are relevant to laser cooling and precision spectroscopy, are shown.\cite{Golub_1988,Takasu_2004,Porsev_2004, Das_2005} The hyperfine structure of the ${}^1P_1$ state is given for the fermionic isotope ${}^{173}\mathrm{Yb}$.}
	\label{fig:levelScheme}
\end{figure}
Yb is a rare-earth element with as many as seven stable isotopes. Its five bosonic isotopes (${}^{168}\mathrm{Yb}$, ${}^{170}\mathrm{Yb}$, ${}^{172}\mathrm{Yb}$, ${}^{174}\mathrm{Yb}$, ${}^{176}\mathrm{Yb}$) are spinless ($I=0$), while the remaining two fermionic isotopes have nuclear spins of $I=1/2$ (${}^{171}\mathrm{Yb}$) and $I=5/2$ (${}^{173}\mathrm{Yb}$). The electronic structure of Yb is dominated by its two valence electrons and hence very similar to alkaline-earth elements. As shown in Fig.~\ref{fig:levelScheme}, its level scheme consists of states with singlet and triplet electronic spin.\cite{Note1} The intercombination transitions connecting singlet and triplet states are only allowed due to imperfect $L$-$S$-coupling and thus are very weak. In particular, the transition from the ground state ${}^1S_0$ to the triplet state ${}^3P_0$ is even forbidden as a single-photon transition for the bosonic isotopes, although small transition amplitudes may be created artificially by magnetically-induced state mixing\cite{Taichenachev_2006} or via multi-photon schemes\cite{Santra_2005, Hong_2005, Ovsiannikov_2007}. In fermionic isotopes, hyperfine interaction yields a non-zero natural linewidth of the transition on the order of $10\ \mathrm{mHz}$. With a lifetime of several seconds\cite{Porsev_2004}, it is metastable on experimentally relevant timescales and gives rise to a series of unique and intriguing features.\cite{Note2}

The ultranarrow ${}^1S_0-{}^3P_0$ transition is not only crucial for novel optical frequency standards\cite{Takamoto_2005, Barber_2006, Poli_2008, Lemke_2009}, its fascinating properties also make it ideal for the spectroscopic analysis of quantum gases. Energy levels of atoms confined to a site of an optical lattice can be probed locally in the optical domain, and due to the coherent nature of the probe even controlled manipulation of quantum states is possible. The wavelength of the lattice acts as a tool to tune the differential AC-Stark shifts of the two states of the spectroscopy transition. In this way, different scenarios can be realized, e.g.~a ``magic'' lattice\cite{Katori_2002} where the main order differential light-shift induced by the trapping potential on the transition is cancelled, thus enabling high-precision spectroscopy. Moreover, state-selective lattices at wavelengths where the polarizability of either ${}^1S_0$ or ${}^3P_0$ vanishes have been proposed\cite{Daley_2008} and allow an unprecedented degree of control of the system.

For laser cooling, two complementary transitions are of interest. The broad principal transition ${}^1S_0\rightarrow{}^1P_1$ allows an extremely strong radiation pressure. In bosonic isotopes, the lack of substructure in the ground state ${}^1S_0$ prevents any Sisyphus cooling on this transition.\cite{Note3} On the other hand, the intercombination transition ${}^1S_0\rightarrow{}^3P_1$ is very narrow and has a \textit{Doppler cooling} limit of merely $T_{D}=4.4\ \mathrm{\mu{}K}$, which provides ideal starting conditions for evaporative cooling. Since the ${}^1S_0$ ground state does not allow magnetic trapping, evaporation and further experimental steps require optical dipole traps. In combination with optical lattices, a setup with wide optical access is therefore desirable for Yb experiments.

\subsection{Side-loaded 2D-MOT}
\begin{figure*}
	\centering
	\includegraphics{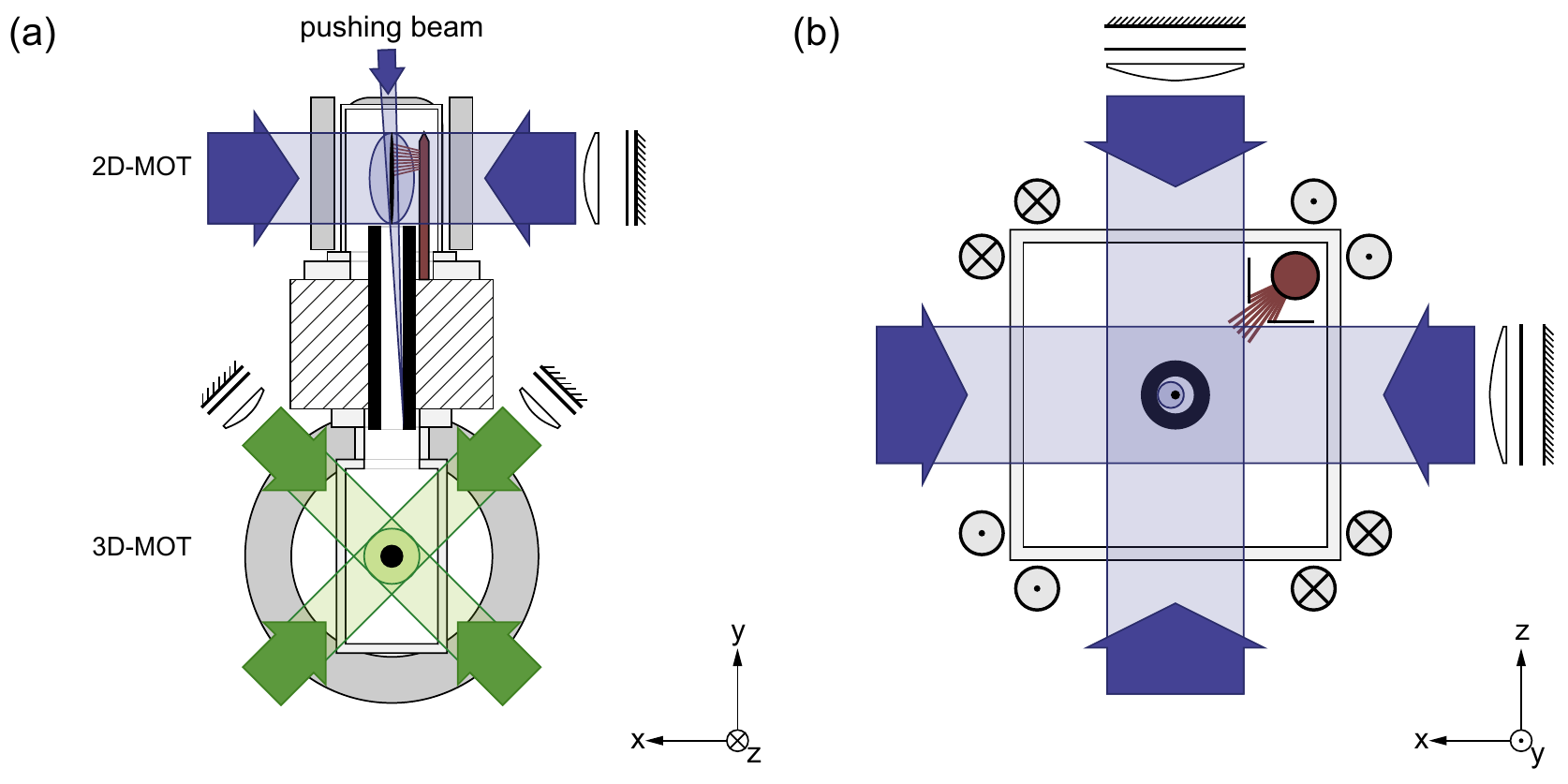}
	\caption{Sketch of the 2D-/3D-MOT system. \textbf{(a)} The 3D-MOT using the intercombination transition ${}^1S_0\rightarrow{}^3P_1$ is loaded from a 2D-MOT operated close to the principal transition ${}^1S_0\rightarrow{}^1P_1$ in a separate glass cell. Both cells are mounted to a central vacuum chamber and connected by a dual differential pumping stage. A pushing beam enhances the loading rate of the 3D-MOT. \textbf{(b)} Top-view of \textbf(a). The 2D-MOT is loaded transversely from the beam of atoms emitted by a dispenser.}
	\label{fig:2d3dsketch}
\end{figure*}
\begin{figure*}
	\centering
	\includegraphics{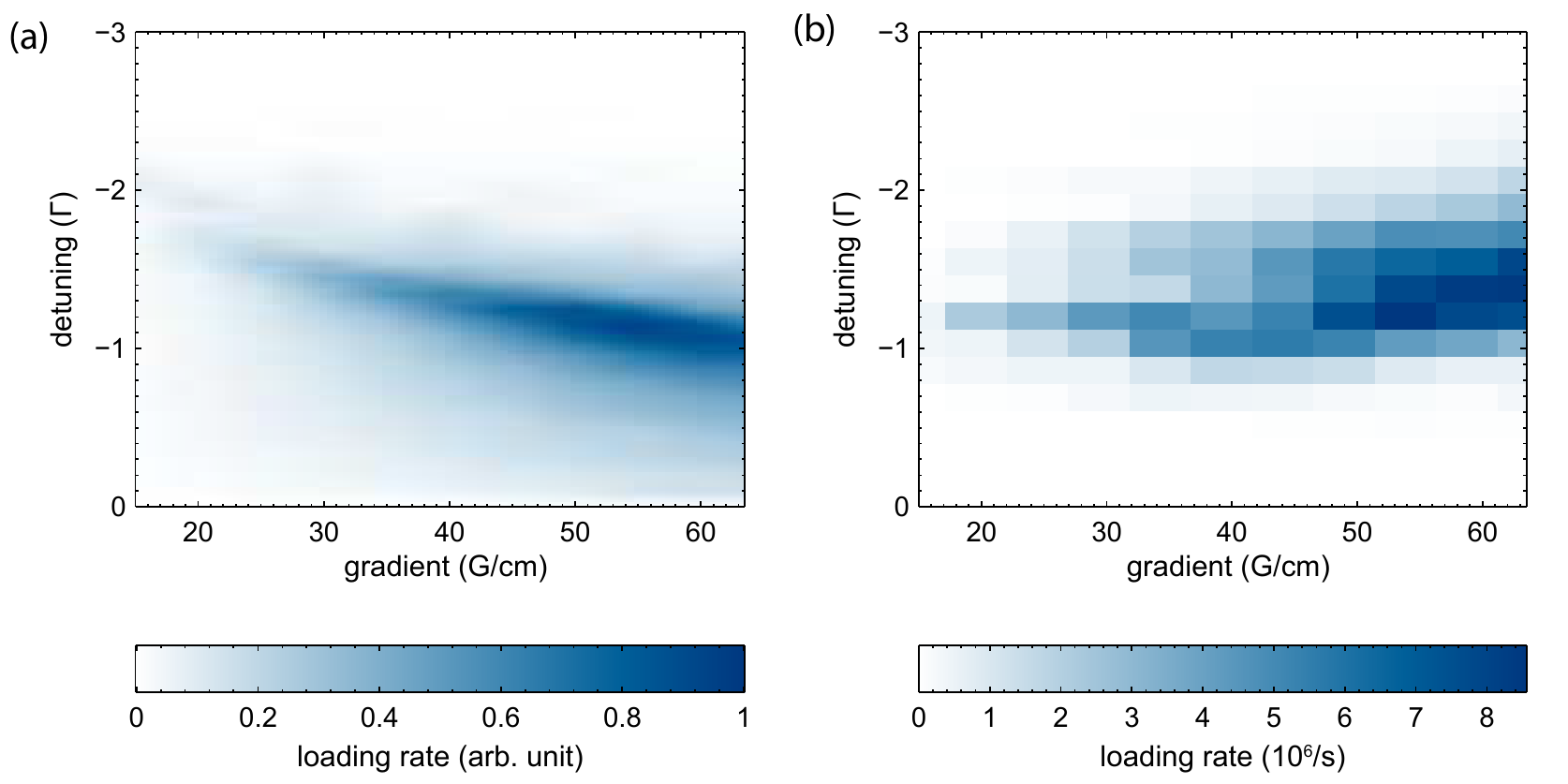}
	\caption{Loading rate of a 3D-MOT of ${}^{174}\mathrm{Yb}$ as a function of magnetic field gradient and detuning of the 2D-MOT. Numerical simulations \textbf{(a)} for a capture velocity of $10\ \mathrm{m}/\mathrm{s}$ of the 3D-MOT and corresponding experimental results \textbf{(b)}. Optimal loading is achieved at a gradient of about $55\ \mathrm{G}/\mathrm{cm}$ and a detuning of $\Delta=-1.2\Gamma$.}
	\label{fig:2dmot_sim_exp}
\end{figure*}
\begin{figure}
	\centering
	\includegraphics{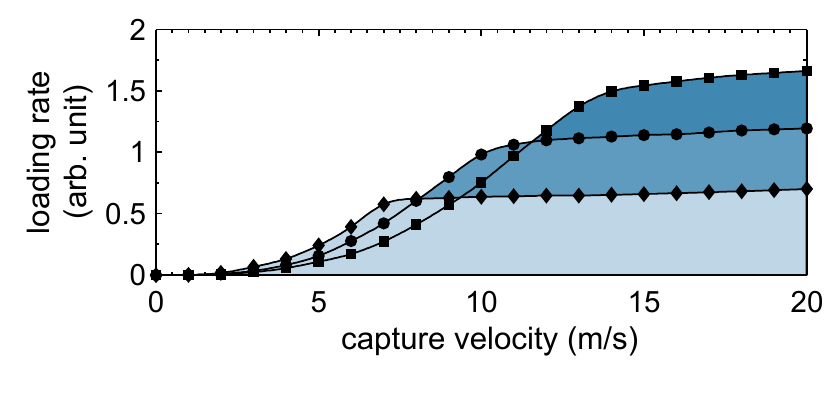}
	\caption{Loading rate of a ${}^{174}\mathrm{Yb}$ 3D-MOT as a function of its capture velocity. Results of numerical simulations are shown for magnetic field gradients of $40\ \mathrm{G}/\mathrm{cm}$ ($\blacklozenge$), $55\ \mathrm{G}/\mathrm{cm}$ ($\bullet$) and $70\ \mathrm{G}/\mathrm{cm}$ ($\blacksquare$) at detunings of $-1.4\Gamma$, $-1.2\Gamma$ and $-1.1\Gamma$, respectively, yielding optimal maximum loading rates.}
	\label{fig:sim_capturevelocity}
\end{figure}
With two nearly closed cooling transitions, Yb is a straightforward candidate for a dual MOT scheme. The principal transition ${}^1S_0\rightarrow{}^1P_1$ is used for cooling in the 2D-MOT to achieve a strong cooling force, while the subsequent 3D-MOT operates on the intercombination transition ${}^1S_0\rightarrow{}^3P_1$ that allows for a final temperature in the microkelvin range. The central concept of our 2D-/3D-MOT setup to overcome the low vapor pressure of Yb is to use a side-loaded 2D-MOT\cite{Tiecke_2009}, but use dispensers instead of an oven. As shown in Fig.~\ref{fig:2d3dsketch}, the 2D-MOT is formed by a perpendicular pair of counterpropagating, circularly polarized laser beams and a 2D magnetic quadrupole field generated by rectangular coils. It is loaded transversely from a dispenser that has been installed within a glass cell. Using the narrow intercombination transition, however, the velocity capture range of the 3D-MOT typically is only a few meters per second. As detailed further in section \ref{section:2d3dmot}, we use laser broadening\cite{Kuwamoto_1999} to enhance the loading rate substantially.

To find the optimal 2D-MOT parameters we have performed detailed numerical simulations. Using the geometry as described above, except for the absence of a pushing beam, we numerically integrated the possible trajectories of Yb atoms emitted from the dispenser for different velocities assuming the mean radiation pressure in the 2D-MOT. We assume that Yb atoms emitted from the dispenser stick to the glass cell at their first contact, which is justified by the high desorption temperature. Thus, atoms perform a ``single pass'' through the 2D-MOT, in strong contrast to 2D-MOTs loading from a thermal background vapor. Spontaneous emissions and collisions have been ignored in this calculation. We sampled the velocity distribution in steps of $1\ \mathrm{m}/\mathrm{s}$ per axis up to a maximum of $250\ \mathrm{m}/\mathrm{s}$ and weighted each trajectory with the flux of the source in the associated velocity bin assuming a thermal Maxwell-Boltzmann distribution. The 2D-MOT beams have an elliptic cross-section of $1\times4\ \mathrm{cm}^2$ in diameter and a homogeneous intensity of $I=0.6I_{sat}$ where $I_{sat}=60\ \mathrm{mW}$ is the saturation intensity of the transition ${}^1S_0\rightarrow{}^1P_1$. From these results we then calculated the rate of atoms entering the 3D-MOT volume at a velocity smaller than the capture velocity. Examples of simulation results for a capture velocity $v_c \approx 10\ \mathrm{m}/\mathrm{s}$ are shown in Fig.~\ref{fig:2dmot_sim_exp}a. Evidently, the 2D-MOT efficiently decelerates and cools a substantial amount of atoms emitted by the dispenser, and the resulting beam of atoms is indeed slow enough to subsequently be captured by the 3D-MOT. The loading rate exhibits a distinct maximum at a magnetic quadrupole field gradient of $55\ \mathrm{G}/\mathrm{cm}$ and a detuning $\Delta_{2D} = -1.2\Gamma$. As expected due to the large linewidth of the cooling transition, the optimal gradient is significantly higher than for typical alkali 2D-MOTs. Experimentally obtained loading rates of our 2D-/3D-MOT system for the same range of magnetic field gradients and detunings are shown in Fig.~{\ref{fig:2dmot_sim_exp}b}, and they are in excellent agreement with the simulation. In particular, the optimal set of parameters is reproduced exactly within the resolution of the scans.

Fig.~{\ref{fig:sim_capturevelocity}} shows the expected loading rates as a function of the capture velocity for several 2D-MOT configurations. It illustrates a strong dependence of the optimal 2D-MOT parameters on the capture velocity around $10\ \mathrm{m}/\mathrm{s}$ and a saturation of the loading rate at values beyond some $20\ \mathrm{m}/\mathrm{s}$.

\section{Experimental setup}
\subsection{Apparatus}
\begin{figure}
	\centering
	\includegraphics{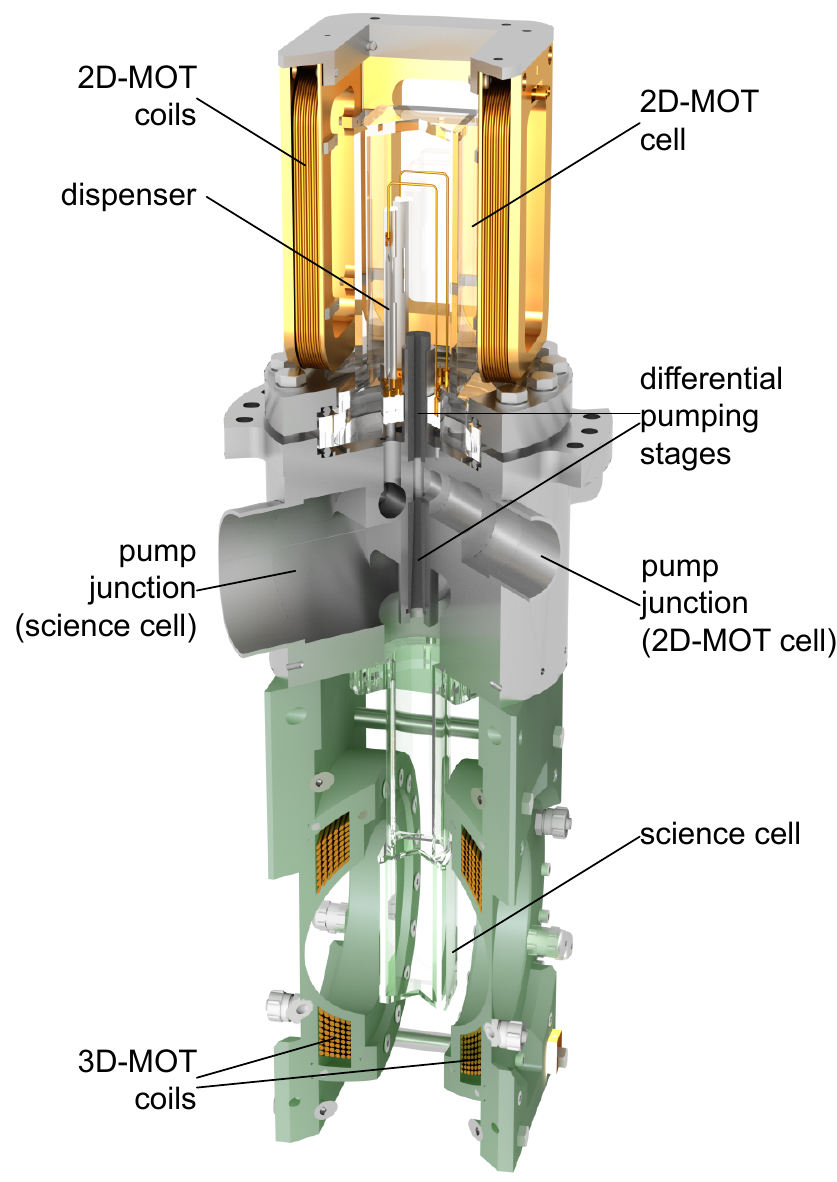}
	\caption{Sketch of the CAD model of the 2D-/3D-MOT apparatus.}
	\label{fig:vacuumSystem}
\end{figure}
The vacuum system of our experimental apparatus consists of two separate glass cells that are mounted to a single stainless steel vacuum chamber. Fig.~{\ref{fig:vacuumSystem}} shows a CAD section view of the apparatus. Both cells are made from synthetic fused silica glass, and broadband anti-reflective coatings have been applied to their outer surfaces. While the 2D-MOT is produced in the upper cell, the 3D-MOT is located in the ``science'' cell below, and all subsequent experimental steps take place in this cell as well.

The central chamber connects the 2D-MOT cell to the science cell via two differential pumping stages, equipped with graphite tubes, along its central axis. These are necessary to maintain the low pressure required to achieve sufficiently long lifetimes in the science cell ($p_{3D}\sim10^{-11}\ \mathrm{mbar}$) even when the 2D-MOT is in operation ($p_{2D}\sim10^{-9}\ \mathrm{mbar}$). Each cell is pumped by a $55\ \mathrm{l}/\mathrm{s}$ ion getter pump, and an additional titanium sublimation pump has been installed for the science cell. The chamber itself has been manufactured from solid, non-ferromagnetic steel with internal tubes to connect the pumps and other equipment to the cells. Its compact designs reduces the distance between both MOTs to $30\ \mathrm{cm}$, which is essential for an efficient transfer of atoms from the 2D-MOT into the 3D-MOT.

The 2D-MOT cell ($50\times{}50\times{}120\ \mathrm{mm}^3$) contains two Alvasource\cite{Note4} AS-4-Yb-500-S Yb dispenser sources that generate the atomic beams loading the 2D-MOT when heated. Both dispensers are connected to a high-power feedthrough and have been mounted upright on an insulating ceramic ring with the emitter slit facing towards the center of the cell. To prevent the Yb atoms from coating any critical surface of the 2D-MOT cell, aluminium plates with rectangular apertures have been installed in front of each dispenser. We typically operate the dispensers at currents of about $6.5\ \mathrm{A}$, which yields a sufficiently large flux of Yb. During operation, the pressure measured close to the 2D-MOT cell rises from some $5\times{}10^{-10}\ \mathrm{mbar}$ to a few $10^{-9}\ \mathrm{mbar}$, likely due to outgassing of the dispensers, as emitted Yb directly sticks to the walls. Two identical pairs of external rectangular coils generate the magnetic quadrupole field for the 2D-MOT. The coils are water-cooled and produce the necessary gradients of more than $50\ \mathrm{G}/\mathrm{cm}$ without radiating a significant amount of heat.

The science cell ($26\times36\times77\ \mathrm{mm}^3$) provides excellent optical access to the atomic samples. A pair of Helmholtz coils generates the magnetic quadrupole field for the 3D-MOT, but it has been designed for large a variety of tasks, such as magnetic traps for another species or bias fields for Feshbach resonances in mixtures. It is capable of producing gradients of up to $200\ \mathrm{G}/\mathrm{cm}$ or homogeneous fields of up to $900\ \mathrm{G}$ and does not reduce the optical access significantly.

\subsection{Optical Setup}
The 2D-MOT is formed by two elliptical beams of about $1\ \mathrm{cm}$ by $4\ \mathrm{cm}$ in diameter that are retroreflected in a cat's eye configuration. We use a commercial diode laser system with a tapered amplifier that is frequency-doubled in a bow-tie cavity and provides about $180\ \mathrm{mW}$ of laser power ($I_0 = 1.9 I_{sat}$) to each beam at a wavelength of $399\ \mathrm{nm}$. The laser is offset-locked to a master ECDL using an additional $\mathrm{GaN}$ laser diode and can easily be scanned across a broad frequency range. The master laser itself is stabilized by fluorescence spectroscopy of an atomic beam. Additional light for pushing and imaging beams is generated by an injection-locked slave laser diode of the same type.

The 3D-MOT has been set up in a retroreflected three-beam-configuration (Fig.~{\ref{fig:2d3dsketch}}). A commercial fiber laser is frequency-doubled in a waveguide and provides $6\ \mathrm{mW}$ ($I_0 = 45 I_{sat}$) and $2 \times 16\ \mathrm{mW}$ ($I_0 = 130 I_{sat}$) for the axial and diagonal beams with a diameter of $15\ \mathrm{mm}$. This laser is stabilized via Doppler-free fluorescence spectroscopy of the atomic beam mentioned above, and its inherent narrow linewidth of less than $50\ \mathrm{kHz}$ is particularly useful for operating a MOT at small detunings.

\section{A 2D-/3D-MOT System for Yb}
\subsection{Basic 2D-/3D-MOT Configuration\label{section:2d3dmot}}
\begin{figure}
	\centering
	\includegraphics{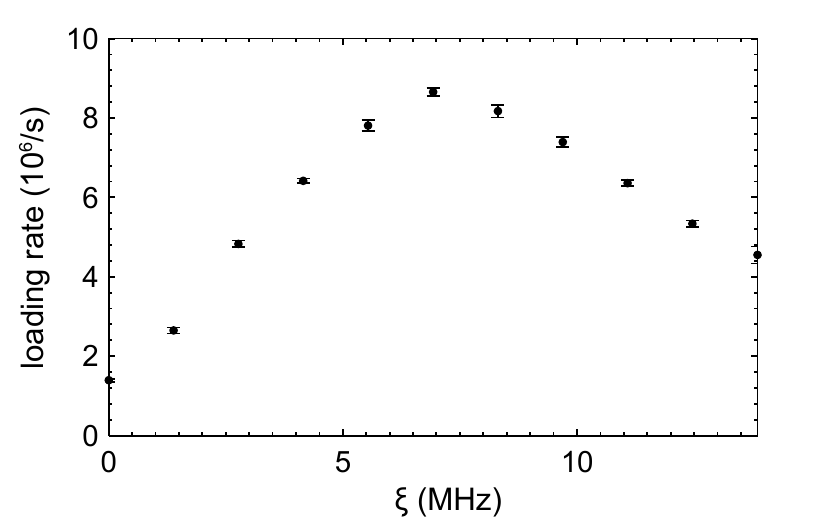}
	\caption{3D-MOT loading rate of ${}^{174}\mathrm{Yb}$ measured as a function of the FWHM $\xi$ of the broadened laser spectrum. The high frequency edge of the spectrum is kept at a constant detuning from the ${}^1S_0\rightarrow{}^3P_1$ resonance frequency.}
	\label{fig:motBroadening}
\end{figure}
Fluorescence in the 2D-MOT cell indicates that atoms are captured in a cigar-shaped region along the axis of the magnetic field. However, the emerging flux of cold atoms cannot be captured efficiently in a single-frequency 3D-MOT setup operated close to the ${}^1S_0\rightarrow{}^3P_1$ intercombination transition, because the narrow linewidth addresses only a very small velocity range. The capture velocity can be increased by actively broadening the frequency spectrum of the laser, as has already been reported for Yb setups using Zeeman slowers\cite{Kuwamoto_1999}. This temporarily introduces additional frequency components, preferably further detuned to the red, at the expense of temperature, as detailed in section \ref{section:motperformance}. Due to the low saturation intensity of the transition, there is typically enough laser power to cover a broad frequency range at a reasonable power spectral density. We use the RF signal driving the AOM that controls the power in the 3D-MOT beams to modulate the laser frequency across several megahertz at a modulation frequency of typically $200\ \mathrm{kHz}$. To keep the minimal red detuning roughly at $-3.2\ \mathrm{MHz}$, we also increase the center detuning accordingly. Fig.~\ref{fig:motBroadening} shows the MOT loading rate as a function of the FWHM $\xi$ of the broadened laser spectrum for the bosonic isotope ${}^{174}\mathrm{Yb}$; we attribute the reduced loading rate above $7\ \mathrm{MHz}$ to insufficient spectral power density.

For optimal loading from the 2D-MOT, we typically operate the 3D-MOT at a center detuning of $\Delta_{3D}=-6.7\ \mathrm{MHz}$, a broadening of $\xi = 7\ \mathrm{MHz}$, and a magnetic field gradient of $2\ \mathrm{G}/\mathrm{cm}$. As shown in Fig.~\ref{fig:2dmot_sim_exp}b, the 2D-MOT is operated at a magnetic field gradient of $55\ \mathrm{G}/\mathrm{cm}$ and a detuning of $\Delta_{2D}=-1.2\Gamma$. Because of the large linewidth of the ${}^1S_0\rightarrow{}^1P_1$ transition used for the 2D-MOT, higher laser power may still increase attainable loading rates significantly, however our experimental setup is currently limited by available laser power and power dissipation in the coils (ca. $400\ \mathrm{W}$). When reducing the laser power for the 2D-MOT by a factor of $0.5$, the loading rate of the 3D-MOT is decreased by a factor of $0.4$.

\subsection{Pushing Beam}
\begin{figure}
	\centering
	\includegraphics{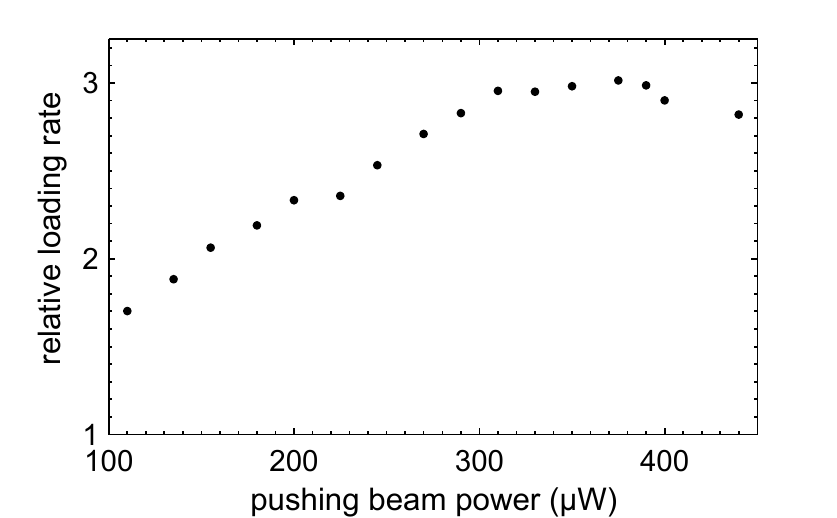}
	\caption{3D-MOT loading rate of ${}^{174}\mathrm{Yb}$ measured as a function of power in the pushing beam. Loading rates are normalized to the value without a pushing beam.}
	\label{fig:pushingBeam}
\end{figure}
The transfer of atoms from the 2D-MOT into the 3D-MOT can be enhanced significantly by a weak additional beam along the axis of the 2D-MOT pushing atoms through the differential pumping stage. We have implemented a single pushing beam as shown schematically in Fig.~{\ref{fig:2d3dsketch}} at a detuning of $\Delta_{p}=-20\ \mathrm{MHz}$ from the ${}^1S_0\rightarrow{}^1P_1$ transition. In Fig.~{\ref{fig:pushingBeam}}, the loading rate of the 3D-MOT is shown as a function of power in the pushing beam. Clearly, a power around $375\ \mathrm{\mu{}W}$ yields an optimal and robust loading rate -- the optimal power depends on its size and the details of its alignment, however, and is given here for reference only. Because of the large linewidth of the ${}^1S_0\rightarrow{}^1P_1$ transition, the radiation pressure of the weak pushing beam exceeds that of the 3D-MOT many times. We therefore align the pushing beam such that it is blocked by the differential pumping tube and does not enter the science cell.

\subsection{MOT Performance\label{section:motperformance}}
For the parameters discussed above, we found a typical value for the 3D-MOT steady-state particle number of $N_{174}^{(0)} = 2\times 10^8$ for $\mathrm{{}^{174}Yb}$, and we typically load $N=6\times{}10^7$ bosonic ${}^{174}\mathrm{Yb}$ atoms within $10\ \mathrm{s}$.

At its optimal loading parameters, the 3D-MOT has a temperature of $T=570\ \mathrm{\mu{}K}$. To reduce the temperature of the sample as close to the Doppler limit of $T_D=4.4\ \mathrm{\mu{}K}$ as possible, the laser-broadening amplitude is gradually reduced to zero within several hundreds of milliseconds, while the center detuning is simultaneously shifted closer to resonance, and the power is reduced to $170\ \mathrm{\mu{}W}$ and $60\ \mathrm{\mu{}W}$ in the diagonal beams and beams along the symmetry axis of the coils, respectively. Furthermore, the MOT is compressed by increasing the field gradient to about $7\ \mathrm{G}/\mathrm{cm}$. With these parameters, the MOT temperature is reduced to typically $T=20\ \mathrm{\mu{}K}$, comparable to temperatures published for other Yb MOTs\cite{Kuwamoto_1999,Hansen_2011}.

\section{Realization of Quantum Degenerate Yb Gases}
\subsection{Crossed Optical Dipole Traps}
For evaporative cooling and trapping of ultracold samples, we have set up a crossed optical dipole trap (ODT) formed by a deep horizontal trap and a vertical trap that creates additional confinement in the crossing region. The horizontal trap is created by a beam of up to $9\ \mathrm{W}$ produced by a solid state laser at $\lambda_{ODT} = 532\ \mathrm{nm}$ and focused down to vertical and horizontal $1/e^2$-waists of $w_y = 18\ \mathrm{\mu{}m}$ and $w_x = 28\ \mathrm{\mu{}m}$, respectively. For the vertical trap, part of the laser power is split off by a partially reflecting mirror and typically up to $1.35\ \mathrm{W}$ are focused to waists of $w_z = 29\ \mathrm{\mu{}m}$ and $w_x = 87\ \mathrm{\mu{}m}$. The powers in each beam are monitored by logarithmically amplified photodiodes and regulated using quartz AOMs that also induce a frequency difference of about $2.5\ \mathrm{MHz}$ between the two beams.

\subsection{Trap Loading}
To transfer atoms into the ODT once the MOT has been loaded, we abruptly switch on the dipole traps and gradually increase the magnetic field gradient to $7\ \mathrm{G}/\mathrm{cm}$ to strongly compress the MOT, while its temperature is reduced as discussed above. Both dipole traps are initially operated at their full powers of $P_h^{(i)} = 9\ \mathrm{W}$ and $P_v^{(i)} = 1.35\ \mathrm{W}$, corresponding to potential depths of $U_h^{(i)}=k_B\times{}625\ \mathrm{\mu{}K}$ and $U_v^{(i)}=k_B\times{}20\ \mathrm{\mu{}K}$ for the horizontal and vertical traps, respectively. Typically $1.2\times{}10^7$ ${}^{174}\mathrm{Yb}$ atoms are trapped in the horizontal ODT at the same temperature measured for the single-frequency MOT. The transfer efficiency strongly depends on the final MOT parameters, in particular the magnetic field gradient.

\subsection{Evaporative Cooling and BEC}
\begin{figure*}
	\centering
	\includegraphics{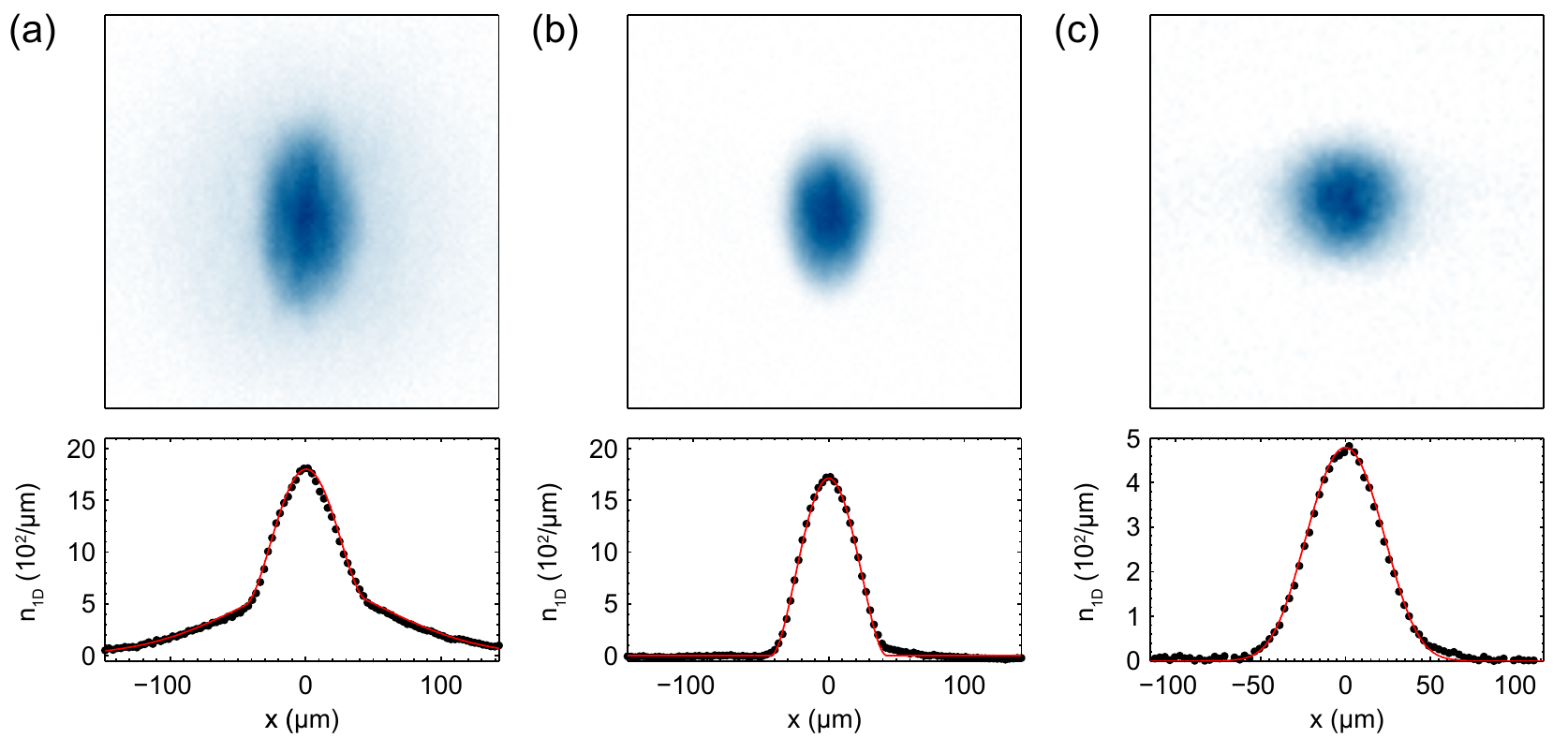}
	\caption{Absorption images of quantum degenerate gases of $\mathrm{Yb}$ taken after time of flight and density profiles (circles) integrated along the vertical axis along with fits of appropriate model functions (red lines). \textbf{(a)} Partially condensed Bose gas of ${}^{174}\mathrm{Yb}$ ($20\ \mathrm{ms}$ TOF). The bimodal fit is composed of a Bose distribution and a Thomas-Fermi density distribution of a harmonic trap potential for the condensate fraction. \textbf{(b)} Nearly pure Bose-Einstein condensate of ${}^{174}\mathrm{Yb}$ ($20\ \mathrm{ms}$ TOF) with a fitted Thomas-Fermi density distribution of a harmonic trap potential. \textbf{(c)} Degenerate Fermi gas of ${}^{173}\mathrm{Yb}$ ($8\ \mathrm{ms}$ TOF). The fit shows the distribution of a non-interacting Fermi gas with $T/T_F=0.32(3)$.}
	\label{fig:bec}
\end{figure*}
Evaporative cooling is performed in two stages by lowering the depths of the ODT. In the first stage, atoms are predominantly evaporated from the horizontal trap, but concentrated into the crossed region of the dipole trap as the temperature decreases. Therefor we exponentially ramp the horizontal trap down to $20\ \mathrm{\mu{}K}$ ($P_h = 0.30\ \mathrm{W}$) within $3.5\ \mathrm{s}$, but keep the vertical trap at constant depth.

Subsequently, atoms are evaporated from the crossed region in a second stage. Within $1.5\ \mathrm{s}$ we exponentially reduce the depth of the horizontal trap to nominally $3\ \mathrm{\mu{}K}$ ($P_h = 40\ \mathrm{mW}$) and relax confinement further by lowering the power of the vertical trap to some $P_v = 35\ \mathrm{mW}$ ($0.5\ \mathrm{\mu{}K}$). It is crucial that the confinement along the axis of the horizontal trap provided by the vertical beam is lowered during the final evaporation stage. However, the vertical trap has virtually no effect on the overall trap depth, and its final power can be varied across a large range to control the resulting trap geometry.

The absorption images in Fig.~\ref{fig:bec}a,b show the emergence of a Bose-Einstein condensate of ${}^{174}\mathrm{Yb}$ during the final stages of the evaporation ramp. We routinely produce nearly pure condensates of $N = 1\times{}10^5$ ${}^{174}\mathrm{Yb}$ atoms with no discernible thermal fraction left. The crossed dipole trap at its final parameters has an effective depth of several tens of nanokelvins and trap frequencies $\omega_{x,y,z}=2\pi\times(118, 188, 56)\ \mathrm{Hz}$ determined from dipole and quadrupole oscillations of the condensate.

\subsection{Degenerate Fermi Gases}
We have produced degenerate Fermi gases of the spin-$5/2$ isotope ${}^{173}\mathrm{Yb}$ using a similar sequence as for ${}^{174}\mathrm{Yb}$, because its $s$-wave scattering length\cite{Kitagawa_2008} is favorable for evaporative cooling in a spin mixture.

The 2D-MOT uses the ${}^1S_0\rightarrow{}^1P_1$ ($F=5/2\rightarrow{}F'=7/2$) hyperfine transition of ${}^{173}\mathrm{Yb}$, but its operation is complicated by the proximity of the $F=5/2\rightarrow{}F'=3/2$ hyperfine transition as shown in Fig.~{\ref{fig:levelScheme}}. This results in a strong reduction of the observed 3D-MOT loading rate at detunings close to the hyperfine splitting, but efficient operation of a 2D-MOT is still possible within a narrow window around the same optimal detuning as observed for bosonic ${}^{174}\mathrm{Yb}$. Even more than in the case of ${}^{174}\mathrm{Yb}$, the loading rate strongly depends on the intensity of the 2D-MOT beams, however.

Due to its lower abundance and reduced 2D-MOT performance, the loading time of the ${}^{173}\mathrm{Yb}$ 3D-MOT is significantly larger than for ${}^{174}\mathrm{Yb}$. With an earlier version, i.e. reduced laser powers in the 2D- and 3D-MOT beams, we obtained the following results: A spin mixture of about $2 \times 10^6$ ${}^{173}\mathrm{Yb}$ atoms was transferred from a MOT of $1.5\times10^7$ atoms to the dipole trap. Thermalization is faster than in ${}^{174}\mathrm{Yb}$ due to the enhanced scattering rate, and we used an evaporation ramp of similar shape as described above, but steeper than for the bosons. After $4\ \mathrm{s}$ of evaporation $3\times{}10^4$ atoms at a temperature of $T/T_F = 0.3$ were left. A typical absorption image of a degenerate Fermi gas is shown in Fig.~\ref{fig:bec}c.

\section{Conclusion}
In conclusion, we have presented a novel scheme for the efficient creation of quantum gases of Yb and similar alkaline-earth-like atoms based on transverse loading of a 2D-MOT. Loading rates observed in the subsequent 3D-MOT are in agreement with numerical simulations of the 2D-MOT setup. The compact 2D-/3D-MOT apparatus facilitates laser cooling of additional species to study quantum-degenerate mixtures and provides excellent optical access to the combined 3D-MOT and science cell. Therefore, it allows a wide range of experiments to be conducted. We routinely produce Bose-Einstein condensates of ${}^{174}\mathrm{Yb}$ and degenerate Fermi gases of ${}^{173}\mathrm{Yb}$ with particle numbers well suited for further experiments.

\begin{acknowledgments}
This work has been supported financially by the Deutsche Forschungsgesellschaft within the SFB 925 and GRK 1355 and the European Union within the FET-Open Scheme (iSense). The authors thank Thomas R{\"u}tzel, Hans Kessler, and Jan Carstens for experimental support.
\end{acknowledgments}

\end{document}